\documentstyle[preprint,aps]{revtex}
\begin{document}
\draft
\title{Temperature dependence of the superconducting gap anisotropy\\
in Bi$_{2}$Sr$_{2}$Ca$_{1}$Cu$_{2}$O$_{8+x}$}
\author{Jian Ma,$^{a}$ C.Quitmann,$^{a}$ R.J.Kelley,$^{a}$\\
H.Berger,$^{b}$ G. Margaritondo,$^{b}$ and M.Onellion$^{a}$}
\address{$^{a}$Department of Physics, University of Wisconsin-Madison,
Madison, WI 53706\\
$^{b}$Institut de Physique Appliqu\'{e}e, Ecole Polytechnique F\'{e}d\'{e}rale,
CH-1015 Lausanne, Switzerland}
\date{Received\ \ \ \ \ \ \ \ \ \ \ }
\maketitle

\begin{abstract}
We present the first detailed data of the momentum-resolved, temperature
dependence of the superconducting gap of  $Bi_{2}Sr_{2}Ca_{1}Cu_{2}O_{8+x}$,
complemented by similar data on the intensity of the photoemission
superconducting condensate spectral area.  The gap anisotropy between the
$\Gamma-\bar{M}$ and
$\Gamma-X$ directions increases markedly with increasing temperature, contrary
to what happens for conventional anisotropic-gap superconductors such as lead.
Specifically, the size of the superconducting gap along the $\Gamma-X$
direction decreases to values indistinguishable from zero at temperatures for
which the gap retains virtually full value along the $\Gamma-\bar{M}$
direction.
\end{abstract}
\pacs{PACS numbers: 73.20.At, 79.60Bm, 74.25.Jb, 74.72.Hs}

The order parameter of high-temperature superconductors is of extreme current
interest, and has been investigated by several techniques, including
penetration depth,\cite{Hardy,Shibauchi} tunneling,\cite{Sun} Raman scattering,
\cite{Devereaux} Josephson
current in an applied magnetic field,\cite{Wollman,Chaudhari} and
angle-resolved photoemission.\cite{Olson,Wells,Shen,Hwu,Kelley}
Angle-resolved photoemission has the advantage of directly investigating
the momentum-dependence of the gap.  Already, results establishing a
marked anisotropy in the
gap at low temperatures have ruled out an isotropic s-wave symmetry order
parameter.\cite{Wells,Shen,Hwu,Kelley}

Our main result is that, {\it contrary to anisotropic-gap conventional
superconductors such as lead,}\cite{Blackford,Blackford2,Blackford3}
{\it the gap anisotropy of
$Bi_{2}Sr_{2}Ca_{1}Cu_{2}O_{8+x}$ increases with increasing
temperature as one approaches the superconducting transition temperature,
$T_{c}$}.  This result was discovered by estimating the gap from
angle-resolved photoemission data, using the BCS-like
lineshape\cite{Chang,Chang2}
and computer code of Olson, Lynch and Liu.\cite{Olson} Our results place
stringent constraints on any theory of high-temperature superconductivity.

Fig.\ \ref{Au-AC_sus} illustrates the experimental and sample standards
we have achieved as prerequisites for our study. Fig.\ \ref{Au-AC_sus}(a)
illustrates an angle-resolved photoemission spectrum of a gold film
deposited {\it in situ}; the temperature of
the film is 36K. The 10-90\% energy width of the Fermi-Dirac distribution
function Fermi edge is 15$\pm$2 meV. Fig.\ \ref{Au-AC_sus}(b)
illustrates magnetic susceptibility
measurements taken for a $Bi_{2}Sr_{2}Ca_{1}Cu_{2}O_{8+x}$ single crystal
sample. The 10-90\% transition temperature
width is 1.3K. To our knowledge, this is the narrowest transition width
published for $Bi_{2}Sr_{2}Ca_{1}Cu_{2}O_{8+x}$ single crystal samples,
and is comparable to the best results on $YBa_{2}Cu_{3}O_{7-x}$
single crystals. Our photoemission measurements were performed
in an ultrahigh vacuum chamber with a base pressure of
$6\ \times\ 10^{-11}$ torr. The light source is the four meter normal
incidence monochromator at the Wisconsin Synchrotron Radiation Center. The
electron energy analyser is a 50 mm VSW hemispherical analyser, mounted on a
two axis goniometer, with an acceptance full angle of 2 degree. The total
energy resolution employed was 25 meV. Samples were
transferred from a load lock chamber and were cleaved {\it in situ} at 35K.
The sample
holder includes the capability to rotate the sample about the surface normal,
at low temperature, for precision alignment with respect to the photon electric
field. The sample crystal structure, and orientation, were determined by {\it
in
situ} low energy electron diffraction (LEED). The sample temperature stability
was $\pm1$K.

We have studied the angular extent of the photoemission superconducting
condensate and its symmetry.\cite{Hwu,Kelley} Here
we concentrate on the temperature dependence of the gap. For this
purpose, we chose two locations in the Brillouin zone where the gap and
the photoemission superconducting condensate spectral area  $n_{s}$ are
large. These points are (a) along the $\Gamma-\bar{M}$
direction(Cu-O-Cu bond
axis in real space) near $\bar{M}$, $k_{x} =\ 0.82\ \AA^{-1}$, and (b) along
the $\Gamma-X$ direction
(Bi-O-Bi direction, without superlattice structure, in real space) at $k_{x}=
k_{y}=\ 0.33\ \AA^{-1}$. For example, pure $d_{x^{2}-y^{2}}$ symmetry
has a maximum
gap along the $\Gamma-M$ direction and zero gap along the $\Gamma-X$ direction.
 Fig.\ \ref{GammaM} illustrates (a) angle-resolved
photoemission data (raw data, without smoothing) taken
at a location along the $\Gamma-\bar{M}$ direction, near
the $\bar{M}$ point, in the
surface Brillouin zone for temperatures from 36K to 95K, and (b) a direct
comparison of the spectra at 35K, 75K and 85K with the normal-state
spectrum at 95K.
The count rate in photoemission superconducting condensate
spectral area at 36K was about 1 kHz. The main temperature-
dependent features of these spectra include: A shift of the leading edge that
reveals the opening of the superconducting gap; a photoemission superconducting
condensate spectral area
with an energy full width at half maximum of 25 meV is observed immediately
below the leading edge; a dip at binding energy of
about 80 meV\cite{Shen,Hwu}
for temperatures up to $0.84T_{c}$. Note that along the $\Gamma-\bar{M}$
direction, the photoemission superconducting condensate spectral area
has been reproducibly observed at temperatures
within 2K of the transition temperature ($T_{c}$) of 83K, and
disappears above $T_{c}$.
Similarly, the shift of the leading edge is quite large, even close
to $T_{c}$. Using
the Olson-Lynch-Liu code,\cite{Olson} we find that the gap opens
quite rapidly below
$T_{c}$, reaching its full value at $0.85T_{c}$.

The results of Fig.\ \ref{GammaM} are in striking contrast to the corresponding
results of Fig.\ \ref{GammaX}, obtained at the location along the $\Gamma-X$
symmetry direction.  The results of Fig.\ \ref{GammaX} include (a) spectra
taken at temperatures from 40K to 95K and (b) a direct comparison of the
spectra at 40K, 75K and 85K with a normal-state spectrum at 95K.
Several differences
compared with Fig.\ \ref{GammaM} are noteworthy. The photoemission
superconducting condensate spectral area is weaker and the
shift of the leading edge of the spectrum at 40K is significantly less than
along the $\Gamma-\bar{M}$ direction. As Fig.\ \ref{GammaX}(b) illustrates,
at temperatures well below $T_{c}$ there is a non-zero superconducting gap.
This is a significant point: it rules out the simplest type of $d$-symmetry
order parameter, specifically pure $d_{x^{2}-y^{2}}$, well below
$T_{c}$.\cite{Chubukov,Monthoux} On the other hand,
the gap becomes indistinguishable from
zero along the $\Gamma-X$ direction at a
temperature (70K) for which the gap is at 90-100\% of full value along the
$\Gamma-\bar{M}$ direction (see Fig.\ \ref{gap-density}, below).

The detailed plot in the superconducting gap with temperature and
symmetry direction is illustrated in Fig.\ \ref{gap-density}(a) for both
the $\Gamma-\bar{M}$ and $\Gamma-X$ directions. We emphasize that the data of
Fig.\ \ref{gap-density}(a) and (b) were obtained on the same sample, and
checked as noted below on additional samples.
Fig.\ \ref{gap-density}(a) illustrates our main result: the gap
anisotropy between
the $\Gamma-M$ and $\Gamma-X$ directions increases with increasing temperature.
For the $\Gamma-M$ direction, the gap is still visible for temperatures as high
as $(0.94-0.98)T_{c}$, and retains its full value up to
$(0.82\pm0.03)T_{c}$.  By contrast, in the $\Gamma-X$ direction the gap
begins to decrease at $(0.57\pm0.03)T_{c}$ and is indistinguishable
from zero at $0.81T_{c}$. Consequently,
the gap anisotropy, already present at low temperatures, increases
as the temperature approaches $T_{c}$. Taking the most
conservative error
bars, the gap anisotropy for the two directions increases from 1.8 at
$0.40T_{c}$ to 14 at $0.85T_{c}$, an increase of at least a factor of 8.

We made several checks to insure that the differences illustrated in
Fig.\ \ref{gap-density}(a) are due to the different symmetry directions.
We observed a temperature dependence of the momentum-resolved photoemission
superconducting condensate spectral area, $n_{s}$.  $n_{s}$ is proportional
to the
number of electrons that are removed by the superconducting transition
from an energy domain roughly equal to the gap, and because of forming
Cooper pairs are found at energies immediately below the gap.
Fig.\ \ref{gap-density}(b)
illustrates the change in $n_{s}$ with temperature, normalized to
the spectral area at $0.40T_{c}$.
We obtained $n_{s}$ by subtracting the normal state(quasiparticle) spectral
area from the superconducting state spectral area between $10-55$ meV binding
energy. We found that the change in $n_{s}$ with temperature was a robust
quantity, not sensitive to exactly what binding energy range was used to
define the photoemission superconducting
condensate spectral area. Although
there is no detailed theoretical calculation of $n_{s}(T)$, several noteworthy
points emerge from the data.
At $0.40T_{c}$, $n_{s}$ is non-zero in both the $\Gamma-M$ and
$\Gamma-X$ directions,
as expected for a non-zero gap in both directions.\cite{Shen,Kelley}
As the temperature increases from the lowest
temperature, $n_{s}$ goes down more
rapidly in the $\Gamma-X$ direction than in the $\Gamma-\bar{M}$
direction; note
particularly the data between $0.50T_{c}-0.80T_{c}$. When the gap in
the $\Gamma-X$ direction
becomes very small (Fig.\ \ref{gap-density}(a)),
$n_{s}$ in the $\Gamma-X$ direction drops to virtually zero.
Further, $n_{s}$ in the
$\Gamma-X$ direction is indistinguishable from zero
for temperatures at which in the $\Gamma-\bar{M}$
direction $n_{s}$ is still appreciable. The  behavior of $n_{s}$ mirrors the
temperature dependence of the gap anisotropy.

As an additional check, we
fabricated samples with different oxygen and cation stoichiometries.  We found
that different samples exhibited the same size gap at low temperatures, one in
the $\Gamma-\bar{M}$ and another in the $\Gamma-X$ direction. We found a
growing
gap anisotropy with increasing temperature$-$ the same behavior as
in Fig.\ \ref{gap-density}(a).

We also checked the effect of the normal state (quasiparticle) binding energy.
We compared samples where the quasiparticle binding energy along
the $\Gamma-\bar{M}$ direction of one sample and along the $\Gamma-X$
direction of a different sample
were the same. We found that the difference illustrated in
Fig.\ \ref{gap-density} persisted. We conclude that the difference illustrated
in Fig.\ \ref{gap-density} is related to the two symmetry directions,
rather than the absolute size of the superconducting gap or the
quasiparticle binding energy from which the photoemission superconducting
condensate arises.

Because quantitative calculations comparing to our experimental data of
Figs.\ \ref{GammaM}-\ \ref{gap-density} are not currently available,
we neither endorse nor rule out specific models.\cite{Chubukov,Monthoux,%
Monthoux2,Mahan,Markiewicz,Combescot,Abrikosov,Klein,Ma,Kotliar,Li,Chakravarty,%
Johnson,Tikofsky,Annett}
Instead, we note how
various models are constrained by our results. Any model must explain
how the gap anisotropy arises.\cite{Markiewicz,Combescot} More stringent is
that
any model must explain how the anisotropy changes from 1.8:1 at $0.40T_{c}$ to
at least 14:1 at $0.85T_{c}$, a change of a factor of 8-9.  The increasing gap
anisotropy as the temperature approaches $T_{c}$ is a peculiar feature
of high-temperature superconductors. A conventional BCS superconductor such as
lead can have a gap anisotropy at low temperatures;\cite{Blackford,%
Blackford2,Blackford3} however, this
anisotropy disappears as $T_{c}$ is approached from below.\cite{Blackford}

Our data can be interpreted in terms of a two-component order
parameter,\cite{Joynt}
of which there are several models.\cite{Kotliar,Li,Chakravarty,Joynt}
One possibility is a model that
exhibits only a $d_{x^{2}-y^{2}}$ symmetry component near $T_{c}$ and
both components at lower temperatures. It is noteworthy that the best fit
to our data at 70K and above along the $\Gamma-X$ direction
yields a zero gap.\cite{Chubukov,Monthoux,Monthoux2}

However, such two-component models\cite{Kotliar,Li,Chakravarty,Joynt}
do not yet
provide a quantitative analysis of the temperature dependence of the two
components. Earlier theoretical work on the superconducting order parameter
symmetry\cite{Annett} indicate that two transition temperatures
should be observed
for a two-component order parameter. We note a recent, unpublished mean
field analysis of a two-component order parameter.\cite{Joseph} The
investigators
in Ref.\ \onlinecite{Joseph} assume an order parameter that is a
mixture of $s-$ and $d-$wave
components. Minimization of the corresponding Ginzburg-Landau free energy
gives a temperature-dependent gap anisotropy in agreement with the data of
Fig.\ \ref{gap-density}(a).

One additional point is clear: our data imply strong-coupling. This is based
on the estimated $2\Delta/kT_{c}$, which is 4.6 for the $\Gamma-\bar{M}$
direction, and has been reported by other authors as exceeding the BCS
value of 3.5.\cite{Gu} Further, the temperature dependence of the gap along the
$\Gamma-\bar{M}$ direction (Fig.\ \ref{gap-density}(a)) is not consistent
with a weak-coupling analysis.\cite{Won}

In summary, we have presented the first detailed, momentum-resolved, study of
the temperature dependence of the superconducting gap and photoemission
superconducting condensate spectral area. We find that the gap anisotropy
increases markedly, by at least a factor of 8-9, as $T_{c}$ is approached.
Within our experimental error, the size of the superconducting
gap decreases to zero along the $\Gamma-X$
direction at a temperature where the gap retains virtually full value along
the $\Gamma-\bar{M}$ direction. Furthermore, neither the estimate of
$2\Delta/kT_{c}$ at low temperatures, nor the temperature dependence of the gap
along the $\Gamma-M$ direction, is consistent with weak-coupling.\cite{Won} The
temperature dependence of the gap anisotropy is mirrored by the temperature
dependence of $n_{s}$. The increase in gap anisotropy with increasing
temperature is related to the two symmetry directions, and appears not to
be due to the absolute size of the gap or the quasiparticle binding energy. The
increasing gap anisotropy with temperature places severe constraints on
all models of high-temperature superconductivity.

We thank David Lynch, Clifford Olson and Rong Liu, who made their computer
code on the superconducting gap available to us. Clifford Olson kindly
showed us his sample heater design. We benefitted from Hong Ding's and
Rong Liu's advice about improving electron energy analyzer resolution.
Michael Winokur kindly showed us the use of
GNUPLOT graphics. Ado Umezawa supervised our AC susceptibility measurements.
We benefitted from conversations with Robert Schrieffer, James Annett,
Andrey Chubukov, Richard Clem, Robert Dynes, Negel Goldenfeld,
Alexander Gurevich, Keith Johnson, Robert Joynt, Davor Pavuna, and David
Tanner.
The staff of the Wisconsin Synchrotron Radiation Center (SRC), particularly
Robert Pedley, were most helpful. Financial support was provided by the U.S.
NSF, both directly(DMR-9214701) and through support of the SRC, by
Ecole Polytechnique F\'{e}d\'{e}rale Lausanne and the Fonds
National Suisse de la Recherche Scientifique, and by Deutsche
Forschungsgemeinshaft.


\begin{figure}
\caption{(a). Gold Fermi edge photoemission spectrum and (b) magnetic
susceptibility data for our $Bi_{2}Sr_{2}Ca_{1}Cu_{1}O_{8+x}$ single crystal.}
\label{Au-AC_sus}
\end{figure}

\begin{figure}
\caption{(a) Angle-resolved photoemission spectra versus temperature for the
($k_{x},k_{y}$) = (0.82,0.0) $\AA^{-1}$ location along the
$\Gamma-\bar{M}$ direction. The photon energy was 21 eV.
The data were taken
from 36K ($0.40T_{c}$) to 95K ($1.15T_{c}$);
(b) direct comparison for the spectra at 36K and 95K, at 75K and 95K, and at
85K
and 95K. The superconducting gap ($\Delta$) obtained for each spectrum is
noted.}
\label{GammaM}
\end{figure}

\begin{figure}
\caption{(a) Angle-resolved photoemission spectra versus temperature for the
($k_{x},k_{y}$) = (0.33,0.33) $\AA^{-1}$ location along the
$\Gamma-X$ direction. A photon energy of 21 eV was used.
The data were taken from 40K ($0.46T_{c}$)
to 95K ($1.15T_{c}$);
(b) direction comparison of the spectra at 40K and 95K, at 75K and 95K, and at
85K and 95K. The superconducting gap ($\Delta$) obtained for each spectrum is
noted. Note that at 75K, the gap is zero, while in Fig. 2(b)
it is non-zero.}
\label{GammaX}
\end{figure}

\begin{figure}
\caption{(a) The size of the superconducting gap for the location in the
$\Gamma-\bar{M}$ direction (diamonds) and in the $\Gamma-X$ direction
(squares) as a
function of temperature. Note that the gap remains at full value up to
$0.85T_{c}$ along the $\Gamma-\bar{M}$ direction, but is reduced starting
at $0.57T_{c}$, and
zero at $0.82T_{c}$, along the $\Gamma-X$ direction; (b) the intensity
$n_{s}$ (see text),
normalized to the value at 40K, versus temperature for the location in the
$\Gamma-\bar{M}$ direction (diamonds) and in the $\Gamma-X$
direction (squares). Note
that $n_{s}$ decreases faster in the $\Gamma-X$ direction, and drops
markedly above
$0.80T_{c}$, compared to the $\Gamma-\bar{M}$ direction.}
\label{gap-density}
\end{figure}

\end{document}